\documentclass[aps, preprint, onecolumn, superscriptaddress]{revtex4}
\usepackage{amsmath,amsfonts,amssymb}
\usepackage{graphicx,amsmath}  
\usepackage{epstopdf}    
\usepackage{relsize}      
\usepackage{color}
\usepackage{MnSymbol}
\usepackage{subfigure}
\def\micron{\:\mu\mbox{m}}

\newcommand{\bea}{\begin{eqnarray}}
\newcommand{\eea}{\end{eqnarray}}

\usepackage{natbib}
\begin{document}

\title{
Discrete Dislocation Dynamics Simulations of Nanoindentation with Pre-stress: Hardness and Statistics of Abrupt Plastic Events}
\author{H. Song$^a$, H. Yavas$^{a}$, E. van der Giessen$^{b}$, S. Papanikolaou$^{a,b,c}$\\
\footnotesize$^a$Department of Mechanical Engineering, Johns Hopkins University, Baltimore, MD, United States.\\
\footnotesize$^b$Department of Applied Physics, Zernike Institute for Advanced Materials, \\
\footnotesize University of Groningen, 9747 AG Groningen, the Netherlands\\
\footnotesize$^c$Department of Mechanical and Aerospace Engineering, \\
\footnotesize West Virginia University, Morgantown, WV, United States.}

\begin{abstract}
{
The yield surface in crystal plasticity can be approached from various directions during mechanical loading. We consider the competition between nanoindentation and tensile loading towards plastic yielding. For this purpose, we develop 
 a two-dimensional discrete dislocation model that is then utilized to investigate the hardness and pop-in event statistics during nanoindentation of single crystal under tensile pre-stress. Indentation is performed by using cylindrical (circular in 2D) indentation with varying radius and under both displacement and load control. Tensile in-plane stress, varying from zero to yield strength, is assigned to investigate the effect of pre-stress on hardness and pop-in statistics. At small indentation depths, the measured hardness is found to be smaller for larger tensile pre-stress; therefore, we conclude that nanoindentation can be used to detect plasticity. When indentation depth is larger, the effect of pre-stress is barely seen. Moreover, we discuss event statistics and the related effect of pre-stress.
}
\end{abstract}

\maketitle

\section{Introduction}

Indentation has been one of the major techniques towards identifying mechanical behavior of a vast list of materials~\citep{oyen2010handbook}. As a technique, it may be labeled as a direct probe, since there is an application of an explicit mechanical force towards identifying a deformation. Moreover, compared with tension/compression tests, nanoindentation is ultra-local and less invasive for the material shape and properties. Over the last few decades, indentation has developed into an important means to interrogate size dependent plasticity.  At the microscale, the indentation size effect has been primarily attributed to the plastic strain gradients that are induced by non-uniform deformation modes~\citep{nix1998indentation}, similar to strain gradients in bending and torsion. These strain gradients have been found to be the essential reason behind the conclusion that ``smaller is harder''~\citep{aifantis1999strain, pharr2010indentation}. Several strain gradient plasticity theories~\citep[e.g.,][]{Hutchinson-SGP, SGP-Gao, Huang2004} have been formulated that can capture the size effects in indentation and many other inhomogeneous
deformation processes~\citep{wei1997steady,huang2000study,particleSGP,3D-rough}. 
However, compression experiments on micro and nanopillars uncovered that size effects are present even in the absence of strain gradients~\citep{Uchic03,Uchic04, Dimiduk2005,Dimiduk2007, Greer2005}. Such size effects are responsible for strengthening at the nanoscale, and are shown to have various origins, such as dislocation source length truncation~\citep{Parthasarathy2007}, dislocation starvation~\citep{Greer2005, Greer2006} and single-arm dislocation sources~\citep{oh2009situ,cui2014theoretical}.

Even though size effects on strength have been the major focus of nanoscale mechanical deformation, it has also become clear that the mechanical response is intermittent and that these too display size effects~\citep{Dimiduk2005,NG2008, Papanikolaou2011, Papanikolaou2012,Papanikolaou2017}. Bursts of plastic deformation~\citep{papanikolaou2017avalanches,sethna2016deformation,maass2017micro} are observed either as stress drops (when imposing displacement control) or strain jumps (when imposing load control). In load controlled indentation, intermittency is also known as ``pop-in"~\citep{Lorenz2003, Bradby2004}, and in recent years, the first pop-in event in relatively pristine crystals~\citep{Durst2006,Shim2008} has been identified to mark the transition from elasticity to plasticity. Following the trend of size effects at the nanoscale, the equivalent stress for the first pop-in is also size dependent: \cite{Morris2011} proposed a statistical model for pop-in during nanoindentation that explained the size-dependence of pop-in stresses in accordance with experimental results. Moreover, \cite{Xia2016} extended pop-in investigations into multiple types of pop-in modes: a primary pop-in with a large displacement excursion and a number of subsequent pop-ins with comparable and small displacement excursions. The distribution of pop-ins in single FCC crystals has also been investigated to
some degree~\citep{wang2012size} for Cu single crystals, finding that the distribution can be fitted by a Gaussian with mean and variance that depend on grain orientation.

It is common in indentation testing that the sample is kept on a stage without any pre-existing applied stresses. However, pre-application of stress on the sample has been used to detect the effect of residual stresses on sample properties~\citep{Suresh1998, Swadener2001, Lee2004, Jang2003}. Early on,  \cite{Tsui1996} and \cite{Bolshakov1996} have revealed the effect of elastic tensile/compressive pre-stress on the measured hardness in microindentation by both experiments and FEM simulations. The increase of tensile pre-stress results in a decrease of the measured hardness up to 6\%, while the effect of compressive pre-stress is barely seen. Moreover, finite element simulations (Bolshakov et al., 1996) suggested that the hardness dependence on pre-stress might be due to the experimental analysis procedures underestimating the indentation contact area. In this paper, we ask the question whether these effects of pre-stress persist for nanoindentation and the pop-in noise?

Commonly, in nanoindentation experiments, due to nanoscale surface roughness, oxide layers, and other defect-like surface features, the regime of depths below 50nm is completely neglected. However, modern nanoindenters in a relatively noiseless environment, only contain 1-2nm depth resolution. Moreover, it is natural to believe that high throughput nanoindentation experiments on a material surface could generate data that may average out extrinsic surface features. Finally, surface polishing techniques have improved to the degree that modern samples may have atomic scale roughness and no other defect-like surface features~\citep{Barnoush2008}. Anyway, it is relatively common that the hardness value as provided by continuous stiffness estimates appears to diverge for depths below 20nm~\citep{li2002review}. Is it possible that this divergence corresponds to a nanomechanical mechanism that emerges below 50nm? In association, it is also quite common that the first pop-ins, when sharp tips are used (corner-cube or Berkovich), appear at a similar scale (5-10nm). Is the first pop-in burst connected to the divergence of hardness below 20nm?  Recently, Bei et al. (2016) described pop-in strength in different regimes: When the stressed volume is free of any pre-existing defects (at very small indentation depth), the pop-in corresponds to the homogeneous nucleation of dislocation at the theoretical strength. At the bulk limit (large indentation depth), the pop-ins are difficult to observe and the deviation from the elastic load–-displacement curves is governed by the bulk yield stress of the material. In the intermediate stage, the pop-in strength shows significant scattering.

There appear to be significant differences for nanoindentation between load and displacement control. Load controlled indentation is normally preferred for the relative simplicity in the experimental setup that does not require any electronic feedback loop~\citep{Vanlandingham2003, Minor2006}. In contrast, nanoindentation discrete dislocation dynamics simulations are most easily performed for displacement control, due to the advantage of not requiring a precise control of the contact area~\citep{Deshpande2005,indentation-Widjaja}. 
Moreover, it is suspected that the hardness and pop-in statistics display differences between load and displacement control since \cite{Cui2016, Cui2017} showed significant differences through 3D discrete dislocation dynamics simulations of nanopillar compression. In that case study, the two different loading modes give different strengths and event (strain burst/stress drops) statistics. Ultimately, it is natural to inquire for analogous differences between various loading modes in nanoindentation.

In this paper, we perform 2D plane-strain discrete dislocation dynamics (2D DDD) simulations  of nanoindentation~\citep{Deshpande2005, widjaja2006effect} to explore and understand the nature of ISEs at the nanometer scale, in relation to the dislocation microstructure. The main application of these simulations is thin crystalline films.  We use 2D DDD to investigate nanoindentation features of a single crystalline sample by both load and displacement controlled protocols. Moreover, in order to identify the sensitivity of ISEs to various dislocation microstructures, constant tensile pre-stress is assigned prior to indentation. This work is focused on an exhaustive statistical exploration of initial conditions in order to unveil the statistical character of the indentation hardness and pop-in effects.  Overall, while 2D-DDD methodology appears restrictive, this is a systematic dislocation plasticity study of nanoindentation that focuses on the effect of pre-stress and different loading modes that should guide precise 3D simulations and also experiments.

The remainder of this paper is organized as follows: Section 2 describes the methodology of our 2D-DDD model for indentation. Section 3 explains the load controlled protocol and presents a comparison with the usual displacement controlled indentation~\citep{indentation-Widjaja}.  Section 4 presents simulation results of indentation hardness at different indentation depths, both for load controlled and displacement controlled protocols. In addition, we focus on the effect of pre-stress on hardness. The effect of pre-stress on the pop-in statistics is discussed in Section 5. In Section 6, we discuss and summarize our results.

\section{Description of the problem}
\label{description}
\begin{figure}[h!] 
 \centering
 \includegraphics[width=0.8\textwidth]{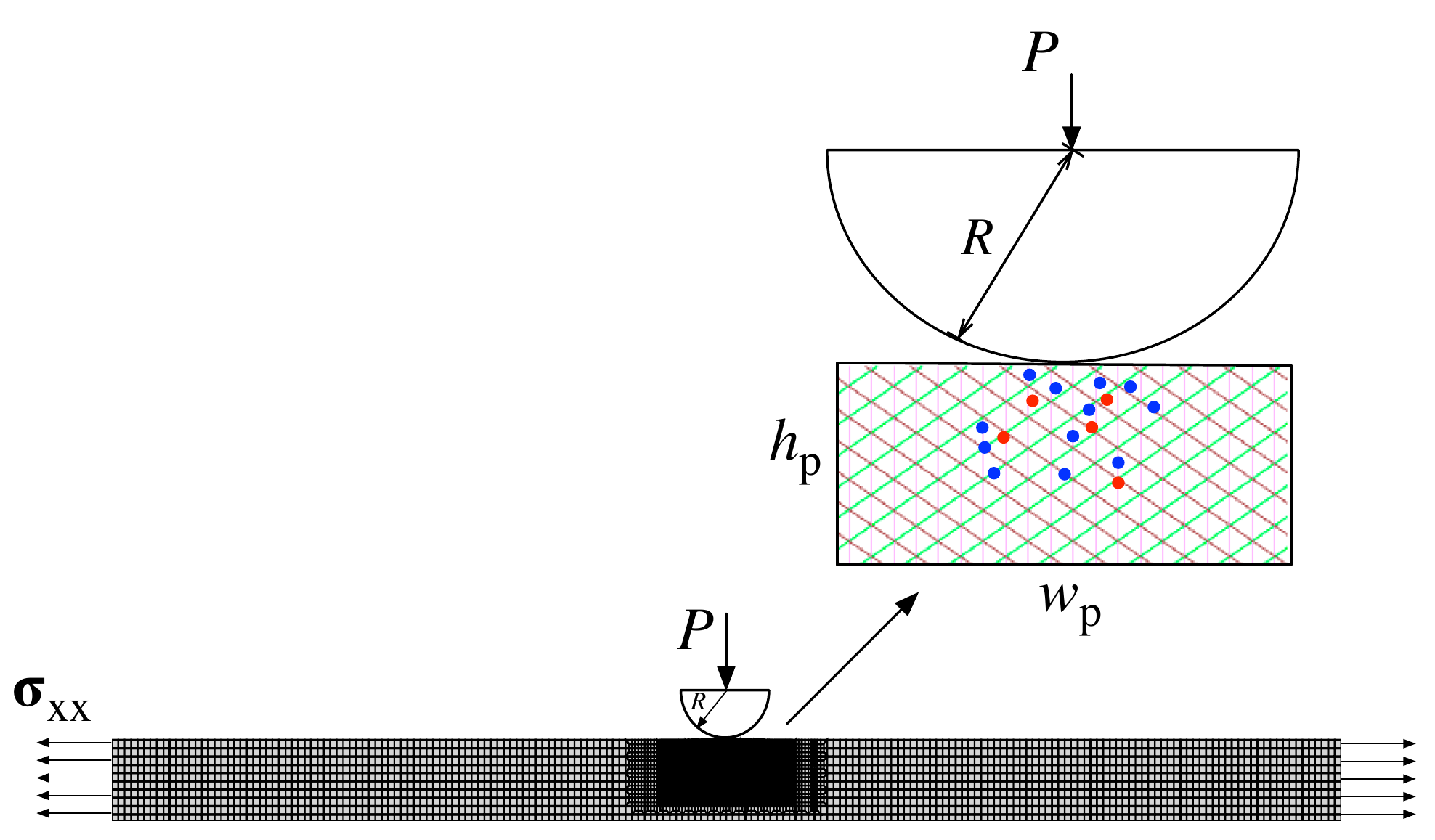}
  \caption{Schematic of indentation problem. The zoom in shows three slip systems (with different colors), red dots stand for dislocation sources while blue dots present dislocation obstacles.}
\label{fig:schematic}
\end{figure}

The model problem is shown in Fig.~\ref{fig:schematic}. A 2D crystal sample with width $1000\mu \rm{m}$ and height $50\mu \rm{m}$ is indented by a rigid circular indenter of radius $R$ at the middle of the sample.  Tensile stress with magnitude $\sigma_{xx}$ is assigned at two ends of the sample prior to indentation. For large enough $\sigma_{xx}$, the system will generate dislocations, and the indentation simulation is started once the total number of dislocations in the sample is relatively stable. 
The finite element mesh is highly refined in the region close to the indenter with the element segment length being 0.005$R$ to accurately capture the evolution of the contact area. Plasticity is limited in a window with size $w_{p}=20\mu \rm{m}$ and $h_{p}=10\mu\rm{m}$ for computational efficiency; once a dislocation reaches the border, the simulation is stopped.  Within the plastic window, three slip systems are considered, with slip directions at $\pm 30^\circ$ and $90^\circ$ respectively, relative to the horizontal direction. Slip planes are spaced at 10$b$ relatively to each other where $b=0.25\rm{nm}$ is the magnitude of the Burgers vector. Dislocation sources (indicated by red dots in Fig.~\ref{fig:schematic}) and obstacles (indicated by blue dots) are randomly distributed on the slip planes. 

Plastic deformation of the crystalline sample is described using the discrete dislocation framework for small strains developed by \citep{vandergiessen1995},
where the determination of the state in the material employs superposition. 
As each dislocation is treated as a singularity in a linear elastic background solid with Young's modulus $E$ and Poisson's ratio $\nu$, 
whose analytical solution is known at any position, this field needs to be corrected by a smooth image field $(\hat{\ })$ to ensure that actual boundary conditions are satisfied. 
Hence, the displacements $u_i$, strains $\varepsilon_{ij}$, and stresses $\sigma_{ij}$ are written as  
\begin{equation}
\label{eq:superposition}
u_i = \tilde{u}_{i}+\hat{u}_{i}, \; \varepsilon_{ij} = \tilde{\varepsilon}_{ij}+\hat{\varepsilon}_{ij}, \;  \sigma_{ij} = \tilde{\sigma}_{ij}+\hat{\sigma}_{ij},
\end{equation}
where the ($\tilde{\ }$) field is the sum of the fields of all $N$ dislocations in their current positions, i.e.
\begin{equation}
 \label{eq:dislocation-field}
\tilde{u}_i=\sum_{J=1}^{N}\tilde{u}_i^{(J)}, \; \tilde{\varepsilon}_{ij}=\sum_{J=1}^{N}\tilde{\varepsilon}_{ij}^{(J)}, \; \tilde{\sigma}_{ij}=\sum_{J=1}^{N}\tilde{\sigma}_{ij}^{(J)}.
\end{equation}
The image fields, indicated by the superposed $\hat{\ }$, are smooth and are obtained by solving a linear elastic boundary value problem using finite elements with the boundary conditions changing as the dislocation structure evolves under the application of mechanical load.

At the beginning of the calculation (before pre-stressing the sample), the crystal is stress and dislocation free. This corresponds to a well-annealed sample, yet with pinned dislocation segments left that can act either as dislocation sources or as obstacles. Dislocations are generated from sources when the resolved shear stress $\tau$ at the source location is sufficiently high ($\tau_{\rm nuc}$) for a sufficiently long time ($t_{\rm nuc}$). 

We only consider bulk sources. The bulk sources are randomly distributed over slip planes at a density $\rho_{\rm{nuc}}$, while their strength is selected randomly from a Gaussian distribution with mean value $\bar{\tau}_{\rm nuc} = 50$ MPa and 10\% standard deviation. The sources are designed to mimic the Frank-Read mechanism in two dimensions~~\citep{vandergiessen1995}, such that they generate a dipole of dislocations at distance $L_{\rm nuc}$, when activated. The initial distance between the two dislocations in the dipole is
\begin{equation}
\label{eq:L_nuc}
L_{\rm nuc}= \frac{E}{4\pi(1-\nu^2)}\frac{b}{\tau_{\rm nuc}},
\end{equation}
at which the shear stress of one dislocation acting on the other is balanced by the local shear stress. Once nucleated, a dislocation can either exit the sample through the traction-free sides, annihilate with a dislocation of opposite sign when their mutual distance is less than $6b$ or become pinned at an obstacle. Point obstacles are included to account for the effect of blocked slip caused by precipitates and forest dislocations on out-of-plane slip systems that are not explicitly described. The strength of the obstacles $\tau_{\rm obs}$ is taken to be $150$ MPa with 20\% standard deviation. They are randomly distributed over the slip planes with a 
density that is eight times the source density~\citep{Papanikolaou2017}.  

Dislocations are considered to move by glide only, driven by the component of the Peach-Koehler force in the slip direction. For the $I$th dislocation, this force is given by
\begin{equation}
\label{eq:P-K}
f^{(I)} = \boldsymbol{n}^{(I)}\cdot\left(\boldsymbol{\hat{\sigma}}+\sum_{J\neq I}{\boldsymbol{\tilde{\sigma}}}^{(J)}\right)\cdot\boldsymbol{b}^{(I)},
\end{equation}
where $\boldsymbol{n}^{(I)}$ is the slip plane normal and $\boldsymbol{b}^{(I)}$ is the Burgers vector of dislocation $I$.
This force will cause the dislocation $I$ to glide, following over-damped dynamics, with an instantanoeus velocity
\begin{equation}
\label{eq:B-v}
v^{(I)} = \frac{f^{(I)}}{B},
\end{equation}
where $B$ is the drag coefficient. In this paper, its value is taken as $B=10^{-4}$Pa s, which is representative for aluminum. A dislocation stays pinned until its Peach-Koehler force exceeds the obstacle-dependent value $\tau_{\rm obs}b$. Each sample contains a random distribution of bulk dislocation sources as well as dislocation obstacles; for each parameter study, we therefore studied 20 realizations. 

The simulation is carried out in an incremental manner, using a time step that is a factor 20 smaller than the nucleation time $t_{\rm nuc}=10\:$ns. At the beginning of every time increment, nucleation, annihilation, pinning at and release from obstacle sites are evaluated. After updating the dislocation structure, the new stress field in the sample is determined, using the finite element method to solve for the image fields~\citep{vandergiessen1995}. 

\section{Nanoindentation with load control protocol}
In displacement controlled indentation, plasticity is characterized by the drop of the indentation force, as seen in previous DD simulations~\citep{balint2006discrete, widjaja2007contact}.  Contact between rigid indenter and crystal is sticky. The contact area is 
computed as the total length of all finite elements that are in contact, like in~\cite{widjaja2007contact}.

In indentation experiments that are typically carried out using load control, plasticity is characterized by displacement bursts (pop-in). A real load controlled indentation simulation needs a contact algorithm to be implemented between indenter and sample. In order to avoid the numerical complexity associated with this, in this paper, we propose a hybrid loading protocol to mimic the load control protocol used in experiments.  The hybrid protocol is essentially displacement controlled based, but with a feedback loop to the indentation rate in order to maintain the indentation force until the indentation force increases. As in load controlled indentation, the position of indenter is adjusted and the indentation force never drops. The assumption in our hybrid protocol is that the indenter can move as fast as needed in order to maintain the load level. In order to capture indentation force drops, the displacement rate is chosen as 0.004 m$\rm{s}^{-1}$ which is much smaller than the one (0.1 m$\rm{s}^{-1}$) used in previous indentation studies~\citep{balint2006discrete, widjaja2007contact}. This is done to avoid the suppression of force drops by very high displacement rates, see Appendix for more details.

\begin{figure}[ht!]
\centering
\subfigure[]{
\includegraphics[width=0.45\textwidth]{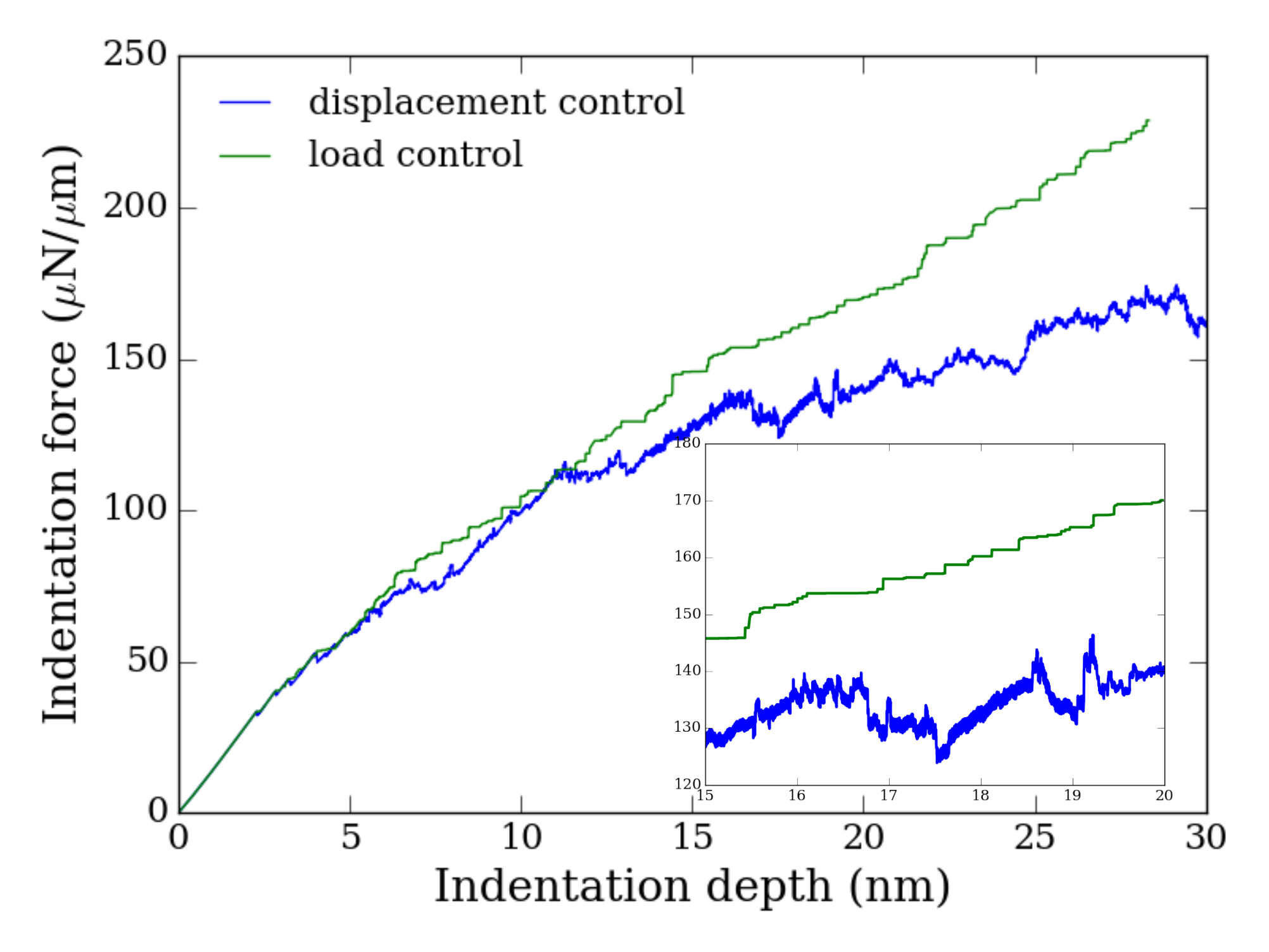}
  }
\subfigure[]{
\includegraphics[width=0.45\textwidth]{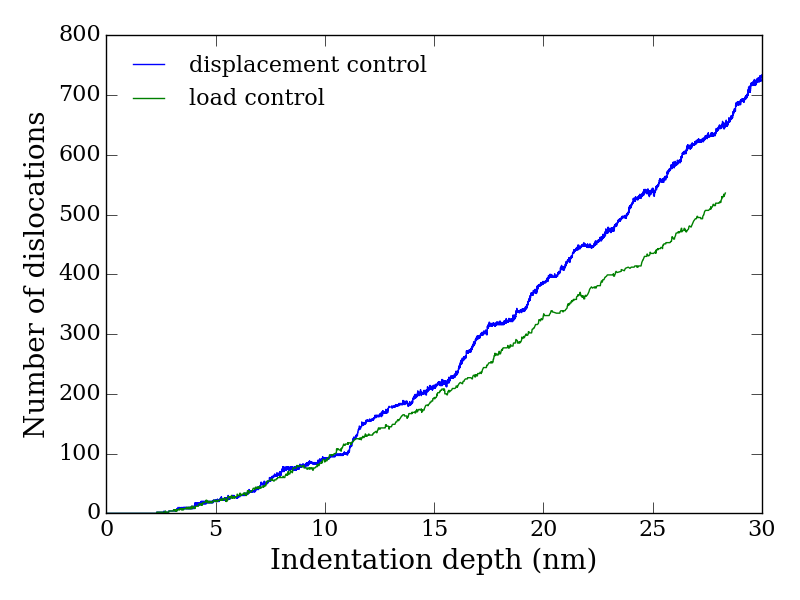}
  }
  \caption{Comparison between load controlled protocol and displacement controlled protocol with dislocation source density $\rho_{\rm nuc}=75/\mu\rm{m}^2$. Two different protocols have the same initial dislocation structure (dislocation sources and obstacles), (a) indentation force versus indentation depth, (b) evolution of the total number of dislocations.}
\label{fig:load-control}
\end{figure}

A comparison between load controlled indentation and displacement controlled indentation is shown in Fig.~\ref{fig:load-control}. 
It can be seen in Fig.~\ref{fig:load-control}(a), using load controlled protocol gradually gives rise to a larger indentation force. The reason is that our load control protocol uses a higher displacement rate when the indentation force tends to drop and that a higher displacement rate yields a higher indentation force. The inset is the zoom in of the curve between indentation depth 15 nm and 20 nm. It can be clearly seen that plasticity is presented as displacement burst under load control while it is force drop under displacement control. The difference on the total number of dislocation can be seen in Fig.~\ref{fig:load-control}(b). At small indentation depth, the two protocols have more or less the same number of dislocations, but due to the changing displacement rate (a higher rate), application of the load control protocol gradually leads to fewer dislocations. 

Details on the effect of the load control protocol are discussed in the Appendix. The different loading protocols are also discussed through nano pillar compression, a problem where the contact area stays constant. 

\section{Dependence of indentation hardness on pre-stress}
\label{hardness-dependence}
\cite{Tsui1996} observed that a tensile pre-stress below the yield strength has a very small influence on the measured hardness in micron-indentation. In this study, we therefore extend tensile pre-stress up to the yield strength to see the influence of pre-stress on measured hardness in nano-indentation. 
The yield strength of the sample with given dislocation parameters ($\rho_{\rm nuc}$, $\rho_{\rm obs}$, $\tau_{\rm nuc}$, $\tau_{\rm obs}$) can be computed by either a force or displacement controlled tensile (or compression) simulation (see Appendix). A simple estimate can be obtained by first noting that for the chosen slip system orientations (cf. Fig. 1), the ratio between the critical resolved shear stress $\tau_{\rm Y}$ and the initial tensile yield stress is 
$\tau_{\rm Y}/ \sigma_{\rm Y} = \frac{1}{2} \rm{sin60}^{\circ}$. 
The critical resolved shear stress is controlled by the weak dislocation sources in the Gaussian distribution with mean value $\bar\tau_{\rm nuc} = 50$ MPa and 10\% standard deviation. Assuming $\tau_{\rm Y}$ to be somewhere between one and a half standard deviation below the mean source strength, we estimate a yield strength of about 100 MPa. This indeed is the value found for the simulations with $\rho_{\rm nuc}=60\micron^{-2}$ in the Appendix. For the lowest source density considered in this study, $\rho_{\rm nuc}=15\micron^{-2}$, the yield strength is likely to be somewhat larger, whereas for $\rho_{\rm nuc}=75\micron^{-2}$ and $150\micron^{-2}$ it may be somewhat lower. Some strain hardening caused by the high density of strong obstacles ensures a stable response up to $\sigma_{\rm Y}= 100$ MPa for the entire range of parameters used. As an additional check we have verified that even for the softest samples, the stress inside the plastic window prior to indentation is equal to the applied pre-stress.

\begin{figure}[h!]
\centering
\includegraphics[width=0.45\textwidth]{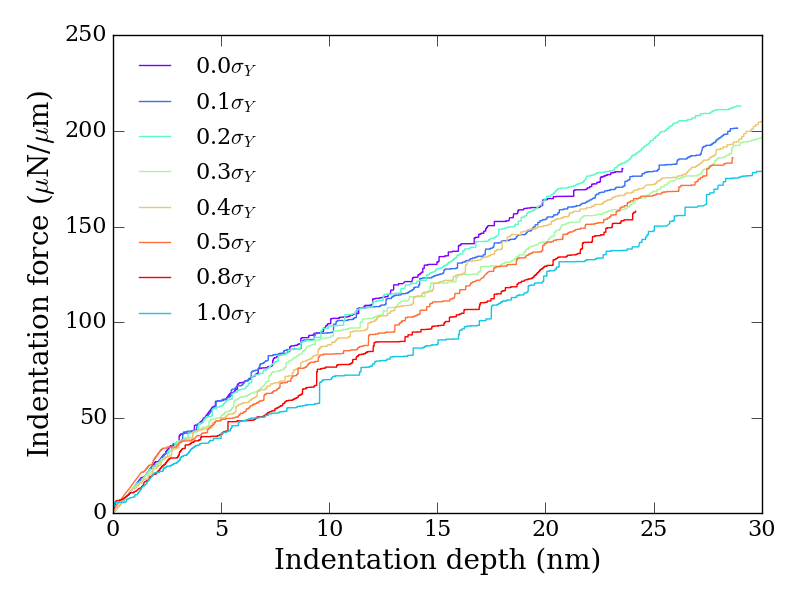}
  \caption{Examples of load versus depth curves obtained under load controlled indentation of a sample with different pre-tensile stresses. $\sigma_{Y}$ is taken as 100 MPa.}
\label{fig:P-h}
\end{figure}

We start indentation using the load control protocol with dislocation source density $\rho_{\rm{nuc}}=75/\mu\rm{m}^2$ (with $\sigma_{\rm{Y}}$=100 MPa). For the same sample (i.e., same dislocation source/obstacle distribution), the indentation force (for indenter radius $R=1\mu\rm{m}$) versus indentation depth ($P-h$ curve) with different pre-stresses $\sigma_{xx}$ is shown in Fig.~\ref{fig:P-h}. We notice that increasing the pre-tensile stress results in a decrease in the indentation force, as one should expect from the enhanced resolved shear stress argument. 

\subsection{Effect of pre-stress on hardness}
\begin{figure}[ht!]
\centering
{
\includegraphics[width=1\textwidth]{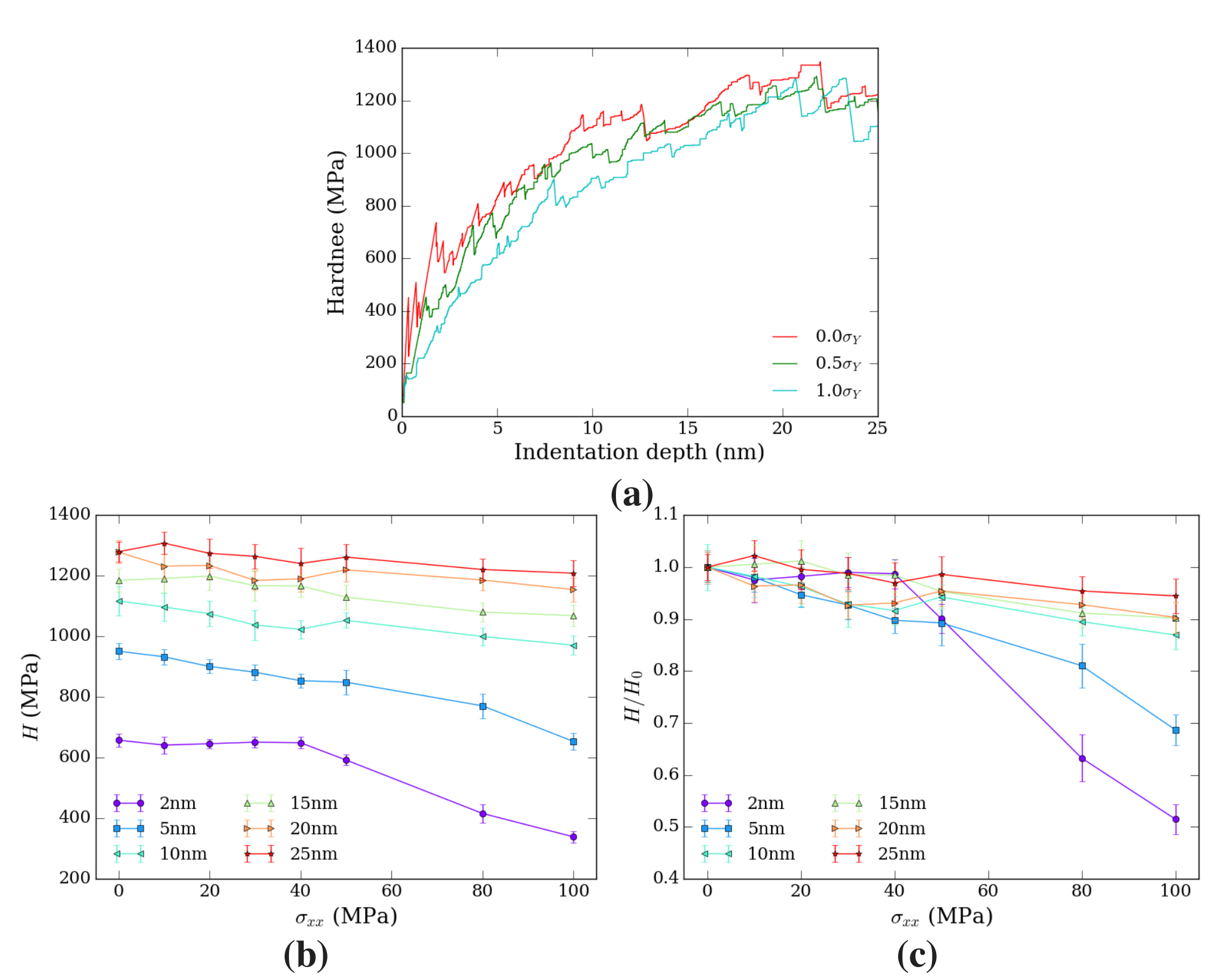}
  }
  \caption{(a) examples of Hardness versus indentation depth for three different tensile pre-stresses obtained by load controlled protocol with dislocation source density $\rho_{\rm{nuc}}=75/\mu\rm{m}^2$. (b) Indentation hardness at different indentation depths for different pre-stresses, (c) hardness normalized by hardness with 0 pre-stress. Error bars represent standard deviation of 20 realizations.}
\label{fig:hardness}
\end{figure}

Indentation hardness is defined as the indentation force $P$ divided by the real contact area $a$ which we calculate following~\cite{widjaja2007contact}. Fig.~\ref{fig:hardness}(a) shows examples of hardness versus indentation depth at various levels of tensile pre-stress. It can be seen that the hardness increases with indentation depth at small indentation depth just as observed without pre-stress~\citep{swadener2002correlation, widjaja2006effect}.
It is important here to note the experimentally confirmed difference between 
a wedge (Berkovich in 3D) and a circular (spherical in 3D) indenter: 
the ISE is controlled by indentation depth in Berkovich indentation whereas it is controlled by the indenter radius in spherical indentation. Moreover, the hardness actually increases with
increasing depth for a spherical indenter. Detailed discussions can be found in~\citep{swadener2002correlation, widjaja2006effect}. 
With different magnitudes of pre-stress, Fig.~\ref{fig:hardness}(b) shows the hardness measured at different indentation depths.   For the same indentation depth, we see that pre-stress reduces the indentation hardness; a small effect at low pre-stress but increasingly more at higher pre-stress levels. In fact, there appears to be a hardness transition at pre-stresses of around 50 MPa (the mean dislocation source strength).  Moreover, the effect of  pre-stress changes with indentation depth, as can be clearly seen in Fig.~\ref{fig:hardness}(c) where the hardness is normalized by the hardness $H_{0}$ with no pre-stress. We see that the hardness decreases with increasing pre-stress at small indentation depth, however, the trend becomes less obvious for larger indentation depth.
\begin{figure}[ht!]
\centering
\includegraphics[width=1\textwidth]{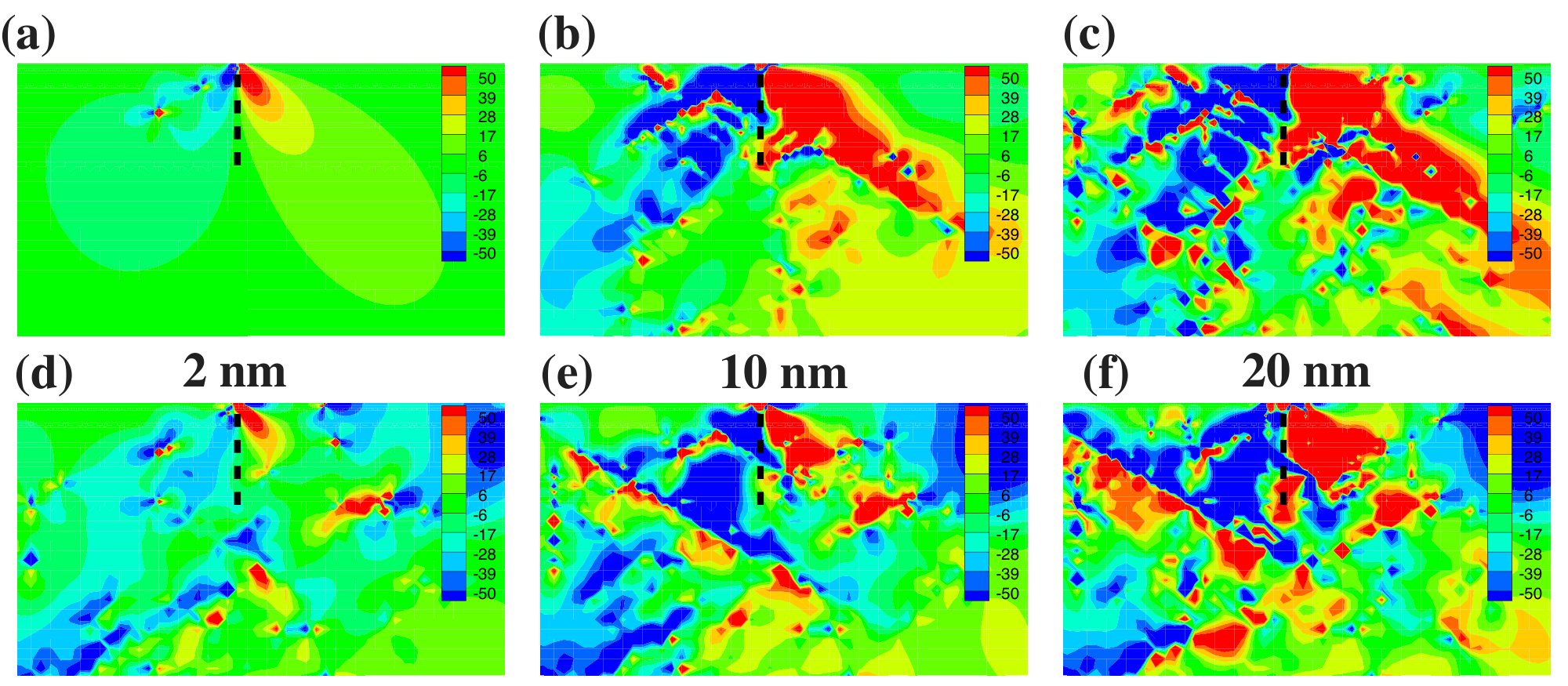}
  \caption{Shear stress filed of indentation under different tensile pre-stresses. (a), (b), (c) are for zero tensile pre-stress, (d), (e), (f) are the results with tensile pre-stress 100 MPa. Contour plots are for different indentation depth as indicated between figures, (a) (d): 2 nm, (b) (e): 10 nm, (c) (f): 20 nm. The thick dashed line indicates the indentation center.}
\label{fig:dislocation-structure}
\end{figure}

The influence of the pre-stress can also be revealed through the stress field under the indenter, shown in Fig.~\ref{fig:dislocation-structure}. Figures~\ref{fig:dislocation-structure}(a), (b), (c) give the stress field at different indentation depths when the tensile pre-stress is zero. We see that with increasing depth, dislocations get nucleated and a high-stress area extends inside the sample. By contrast, when the tensile pre-stress is 100 MPa, shown in Fig.~\ref{fig:dislocation-structure}(d), (e), (f), a high-stress area has been induced by tension (because of nucleated dislocations). Increasing the indentation depth changes the stress field, but not as much as in the case of zero pre-stress.

\subsection{Effect of dislocation source density}
\label{densityeffect}
\begin{figure}[ht!]
\centering
\includegraphics[width=0.9\textwidth]{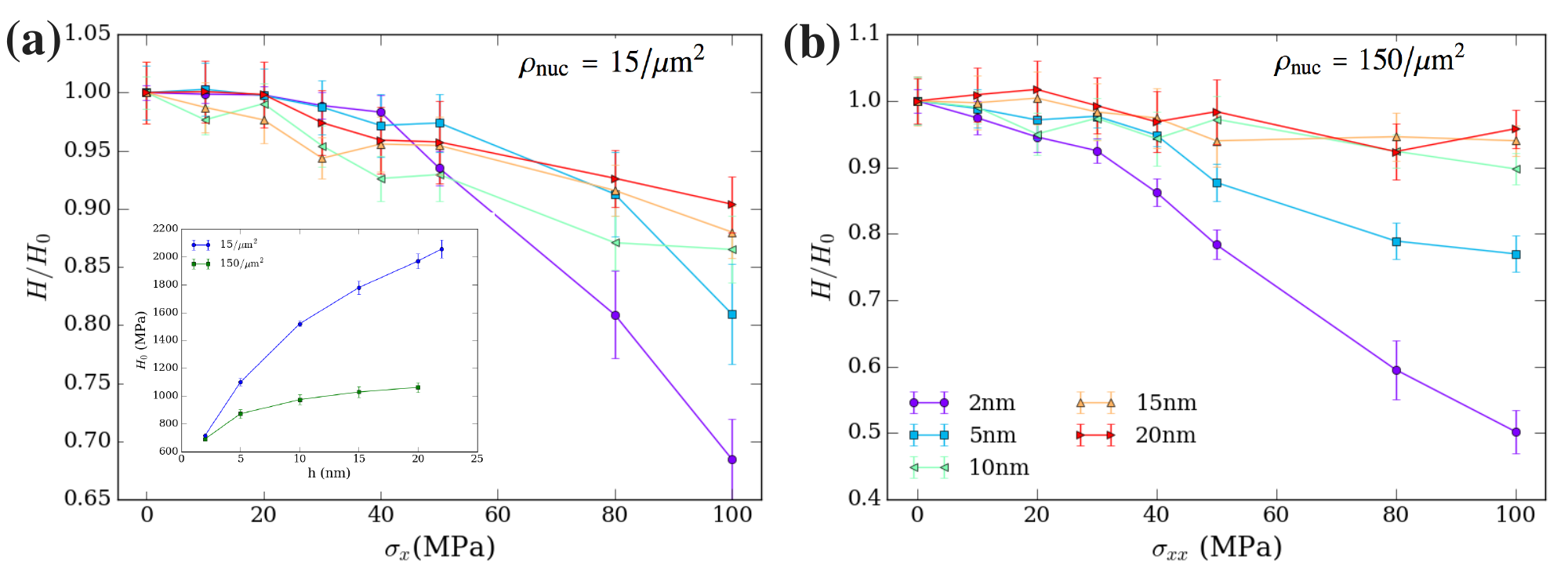}
  \caption{Normalized hardness (by hardness with 0 pre-stress) versus tensile pre-stress obtained by load controlled protocol with different dislocation source densities. (a) $\rho_{\rm{nuc}}=15/\mu\rm{m}^2$, (b) $\rho_{\rm{nuc}}=150/\mu\rm{m}^2$. The inset in (a) is the hardness with zero pre-stress at different depth for two different dislocation source density. Figure (a) and (b) share the same legend  shown in (b). }
\label{fig:density}
\end{figure}

It has been reported in many previous 2D DDD studies of different problems~
\citep{widjaja2006effect, Papanikolaou2017} that the density of dislocation sources can restrict plasticity if the system does not have a surplus of dislocation sources.  The computations at a source density $\rho_{\rm{nuc}}=75/\mu\rm{m}^2$ reported above,  reveal an effect of pre-stress on hardness. In this section, we report results for different dislocation source densities: $\rho_{\rm{nuc}}=15/\mu\rm{m}^2$ and $\rho_{\rm{nuc}}=150/\mu\rm{m}^2$. It is seen in the inset of Fig.~\ref{fig:density}(a) that the lower source density gives rise to the larger hardness at the same indentation depth, just like in the previous 2D DDD indentation study by \cite{widjaja2006effect}. Moreover, it is worthy to note that for the lower source density (Fig.~\ref{fig:density}(a)), the effect of pre-stress on hardness is still present at an indentation depth of 20 nm while the effect disappears for the highest source density (Fig.~\ref{fig:density}(b)).  

\subsection{Effect of indenter radius}
\label{radius}
\begin{figure}[h!]
\centering
\includegraphics[width=0.9\textwidth]{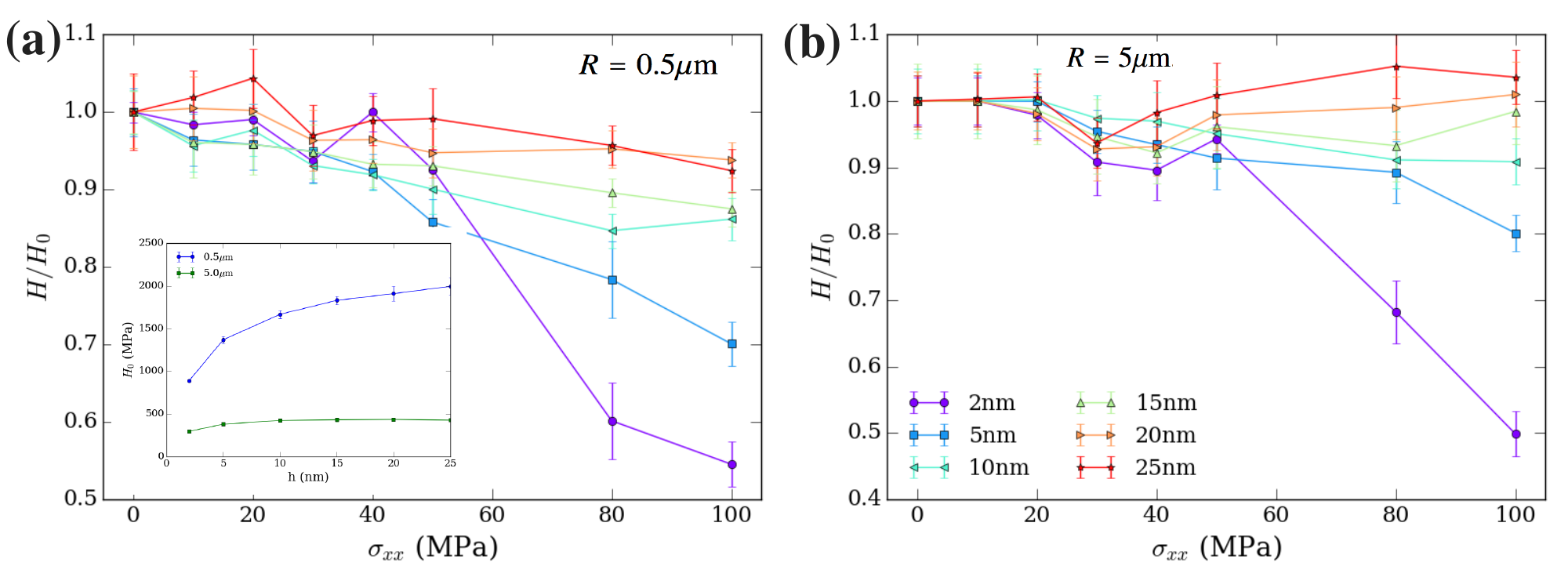}
 \caption{Normalized hardness (by hardness with 0 pre-stress) versus tensile pre-stress obtained by load controlled protocol with dislocation source density $\rho_{\rm{nuc}}=75/\mu\rm{m}^2$, but different indenter radii. (a) $R=0.5\mu\rm{m}$, (b) $R=5\mu\rm{m}$. 
 The inset in (a) is the hardness with zero pre-stress at different depth for two different radii. Figure (a) and (b) share the same legend  shown in (b).}
\label{fig:radius}
\end{figure}

The results reported so far are for an indenter radius $R=1\mu\rm{m}$; in Fig.~\ref{fig:radius} we present the results for a smaller indenter $R=0.5\mu\rm{m}$ and for a larger indenter $R=5\mu\rm{m}$. It can be seen in the inset of Fig.~\ref{fig:radius}(a) that at the same indentation depth, the smaller indenter gives rise to the larger hardness. This is consistent with the typical spherical (circular) indentation size effect~\citep{swadener2002correlation,widjaja2006effect}. Furthermore, we see that for a larger indenter, the hardness transition (decreasing hardness with increasing tensile pre-stress) disappears at a smaller indentation depth (compare Fig.~\ref{fig:radius}(a) and Fig.~\ref{fig:radius}(b)).  

\subsection{Comparison with displacement control}
\begin{figure}[ht!]
\centering
\subfigure[]{
\includegraphics[width=0.45\textwidth]{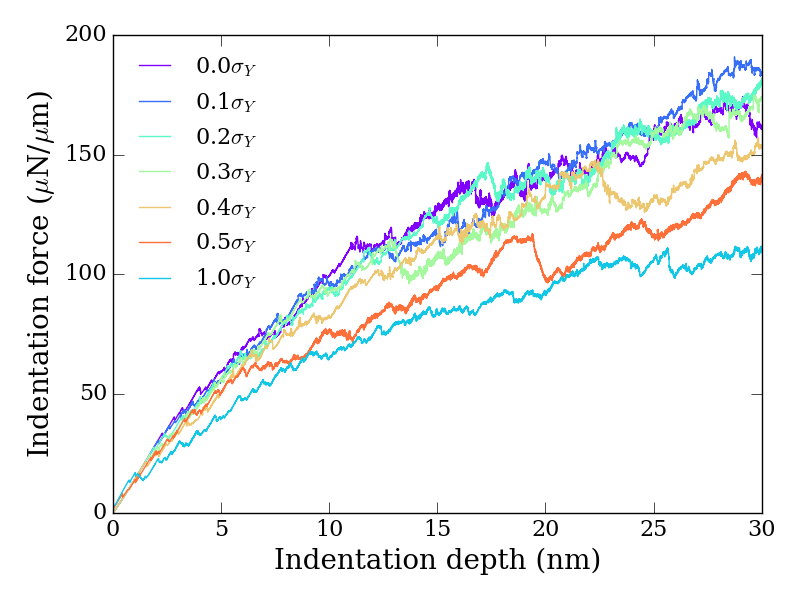}
  }
\subfigure[]{
\includegraphics[width=0.45\textwidth]{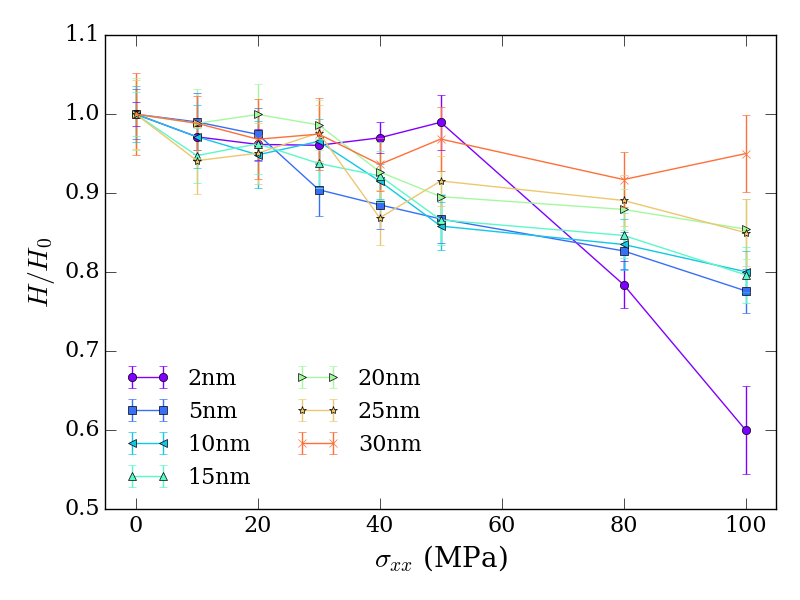}
  }
  \caption{(a) Indentation force versus depth under displacement control with dislocation source density $\rho_{\rm{nuc}}=75/\mu\rm{m}^2$, (b) effect of pre-stress on normalized hardness. $\sigma_{\rm Y}$ is taken as 100MPa.}
\label{fig:displacement-curves}
\end{figure}
Fig.~\ref{fig:displacement-curves}(a) shows the indentation force versus depth for displacement controlled indentation under different pre-stresses.  
It is worthy to note that under displacement controlled indentation, plasticity is characterized by a drop of indentation force, while under load control, it is characterized by a jump (or burst) of the indentation depth.  
Even though the difference in loading mode leads to the difference of indentation force depth curves, the effect of pre-stress on hardness is similar, see Fig.~\ref{fig:displacement-curves}(b). 

\section{Pop-in statistics}
While the emphasis so far has been on the effect of pre-stress on the indentation hardness, in this section, we try to unveil the effect of pre-stress through the statistics of those signals. 

In annealed crystals, the burst of indentation depth appearing during load control is normally called pop-in, and the first pop-in relates to the nucleation of dislocations~\citep{Morris2011}. Using 3D DDD, \cite{fivel-indentation} related pop-in to the fast multiplication of dislocations below the indenter. When the dislocation network formed by fast dislocation multiplication is broken by indentation, pop-in takes place.  In our model, pre-stress can induce dislocations prior to indentation, in which case the burst is caused by those pre-stress induced dislocations, not by dislocation nucleation. 

\begin{figure}[ht!]
\centering
\includegraphics[width=0.9\textwidth]{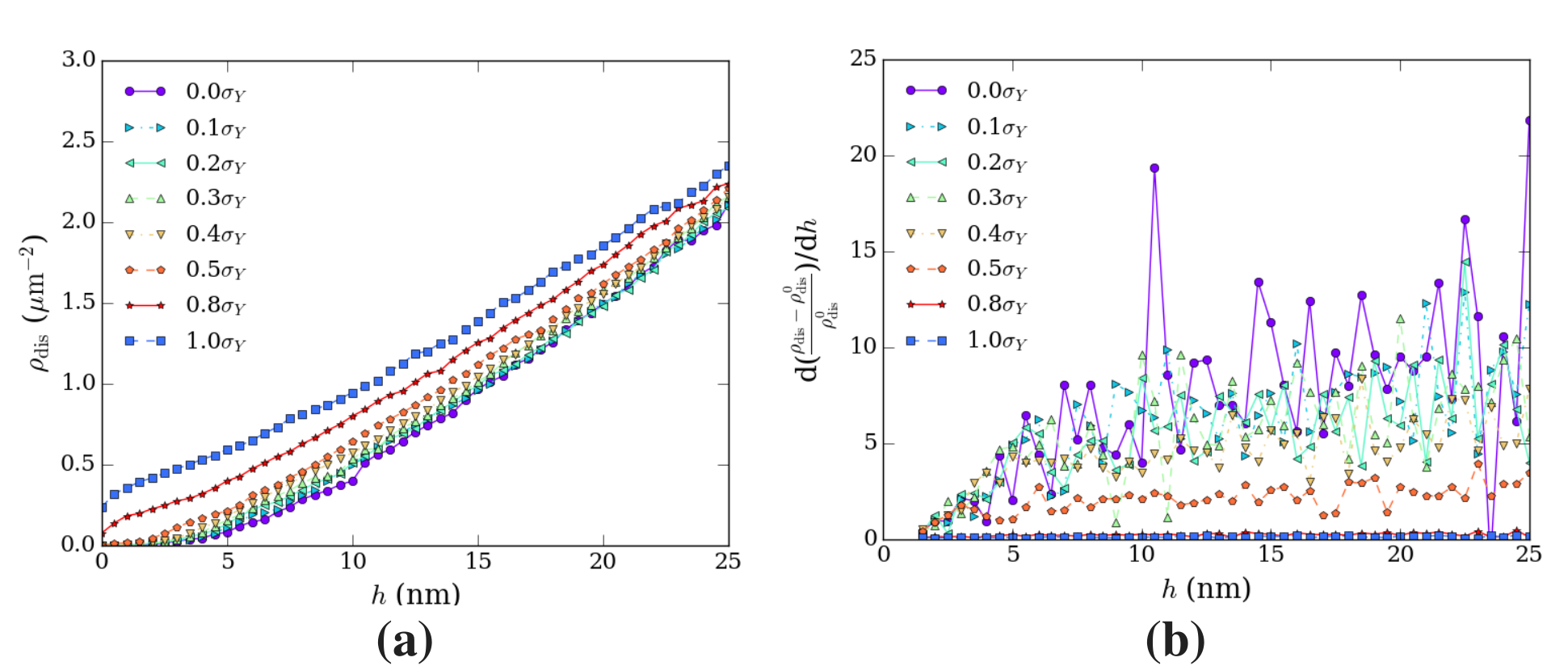}
  \caption{(a) Evolution of dislocation density. (b) change of total number of dislocations. $\rho_{\rm dis}^{0}$ is the initial dislocation density. For large pre-stress, $\rho_{\rm dis}^{0}$ is determined by the pre-stress, for small pre-stress (no pre-stress induced dislocations), $\rho_{\rm dis}^{0}=0.01/\mu\rm{m}^2$. Legend indicates the magnitude of the pre-stress where $\sigma_{\rm Y}=100$MPa.}
\label{fig:dislocation-density}
\end{figure}

Fig.~\ref{fig:dislocation-density}(a) shows the average dislocation density $\rho_{\rm dis}$ over the entire plastic window (200$\mu\rm{m}^2$) at different indentation depths. It is seen that larger pre-stress introduces some dislocations, but the dislocation density becomes almost the same at larger indentation depth irrespective of the pre-stress level.  In order to see the connection between  pop-ins and the nucleation of dislocations, we plot the rate of change of the dislocation density during indentation in Fig.~\ref{fig:dislocation-density}(b). It can be clearly seen that at low pre-stress levels, the growth rate of dislocation density is higher, which means that dislocation nucleation is dominant. At high pre-stresses, plasticity is mainly dominated by pre-stress induced dislocations.

\begin{figure}[ht!]
\centering
\includegraphics[width=0.45\textwidth]{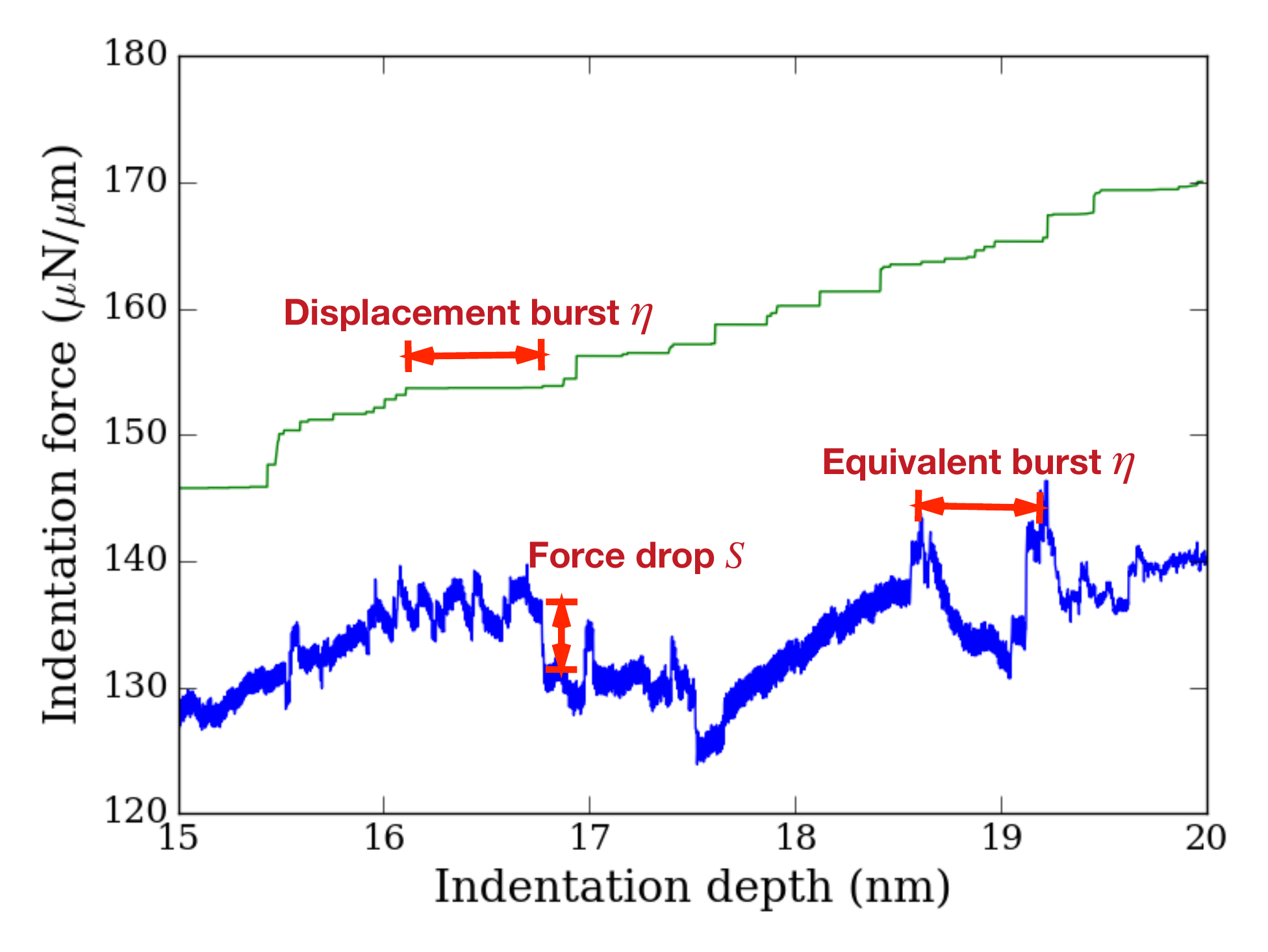}
  \caption{Definition of event: displacement burst in load controlled indentation, force drop and equivalent displacement burst in displacement controlled indentation.}
\label{fig:pop-in-definition}
\end{figure}

The most straightforward way to unveil pop-in statistics is to analyze the displacement burst or the force drop in the indentation force-depth curve. The definition of a pop-in event is indicated in Fig.~\ref{fig:pop-in-definition} on the response curves shown in the inset of Fig.~\ref{fig:load-control}(a). 

\begin{figure}[b!]
\centering
\includegraphics[width=0.9\textwidth]{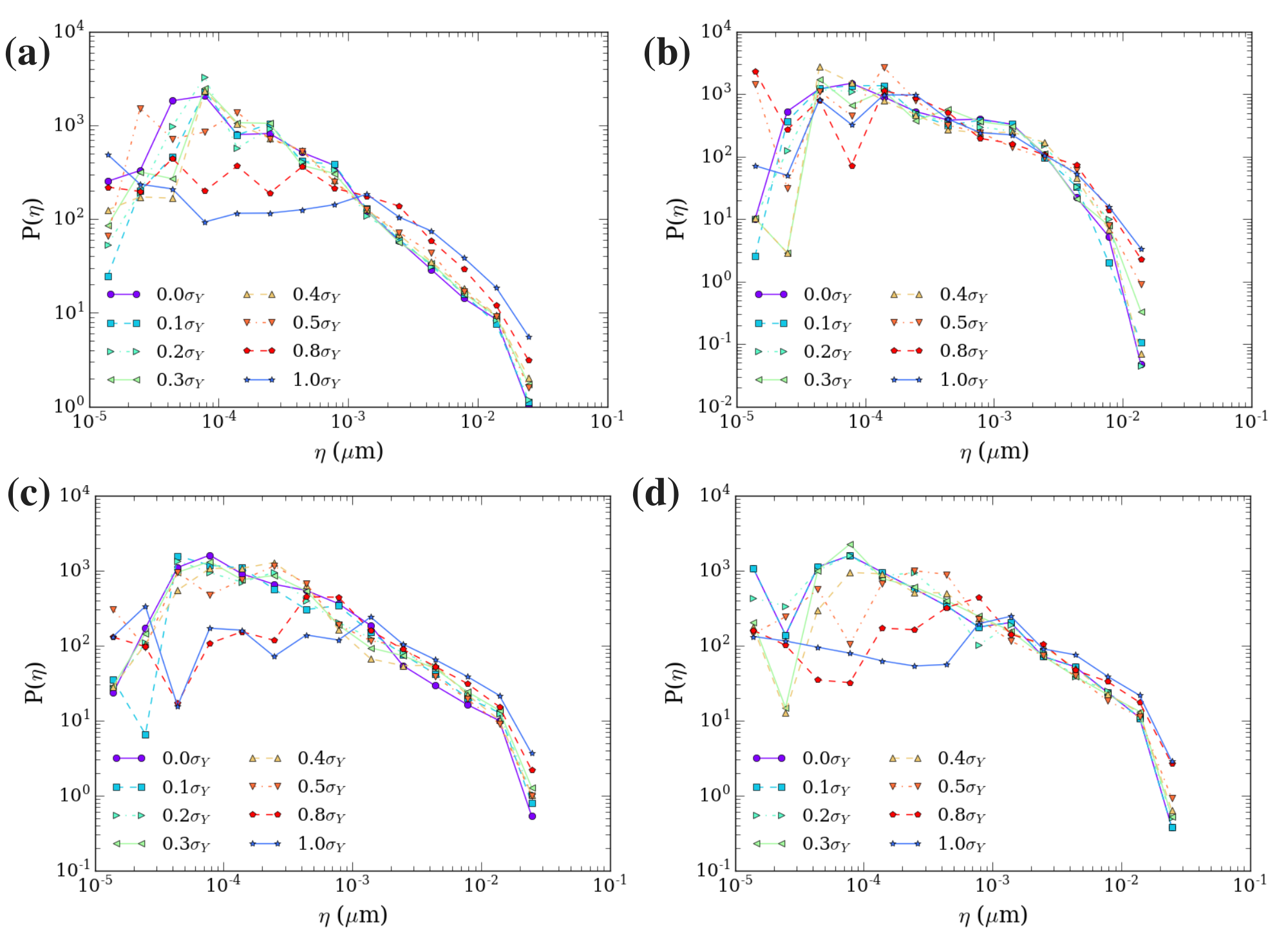}
  \caption{Probability event distribution $P(\eta)$ as function of event size $\eta$ under different pre-stresses for different material properties and indenter size (a) $\rho_{\rm{nuc}}=150/\mu\rm{m}^2$,  $R=1\mu \rm{m}$; (b) $\rho_{\rm{nuc}}=15/\mu\rm{m}^2$, 
$R=1\mu \rm{m}$; (c) $\rho_{\rm{nuc}}=75/\mu\rm{m}^2$, $R=0.5 \mu \rm{m}$; (d) $\rho_{\rm{nuc}}=75/\mu\rm{m}^2$, $R=5\mu \rm{m}$. $\sigma_{Y}$ is taken as 100 MPa.}
\label{fig:pop-in-statistics}
\end{figure}

Fig.~\ref{fig:pop-in-statistics} shows statistics of pop-in of samples under load control. Here, $\eta$ is the the magnitude of a single indentation depth burst, while $P(\eta)$ is the probability density. 
We see that for small pre-stress (up to 0.4$\sigma_{\rm Y}$, i.e., 40 MPa), pop-in has power law statistics spanning two orders of magnitude, except for very small $\eta$,.  However, at a pre-stress $\ge 50$ MPa (i.e., the average dislocation source strength), dislocations induced by pre-stress yield higher probability for very small $\eta$ and at the same time,  they give rise to higher probability for bigger $\eta$. This means that pre-stress induced dislocations not only introduce noise in statistics, but also contribute to big plasticity events.

\begin{figure}[ht!]
\centering
\includegraphics[width=0.9\textwidth]{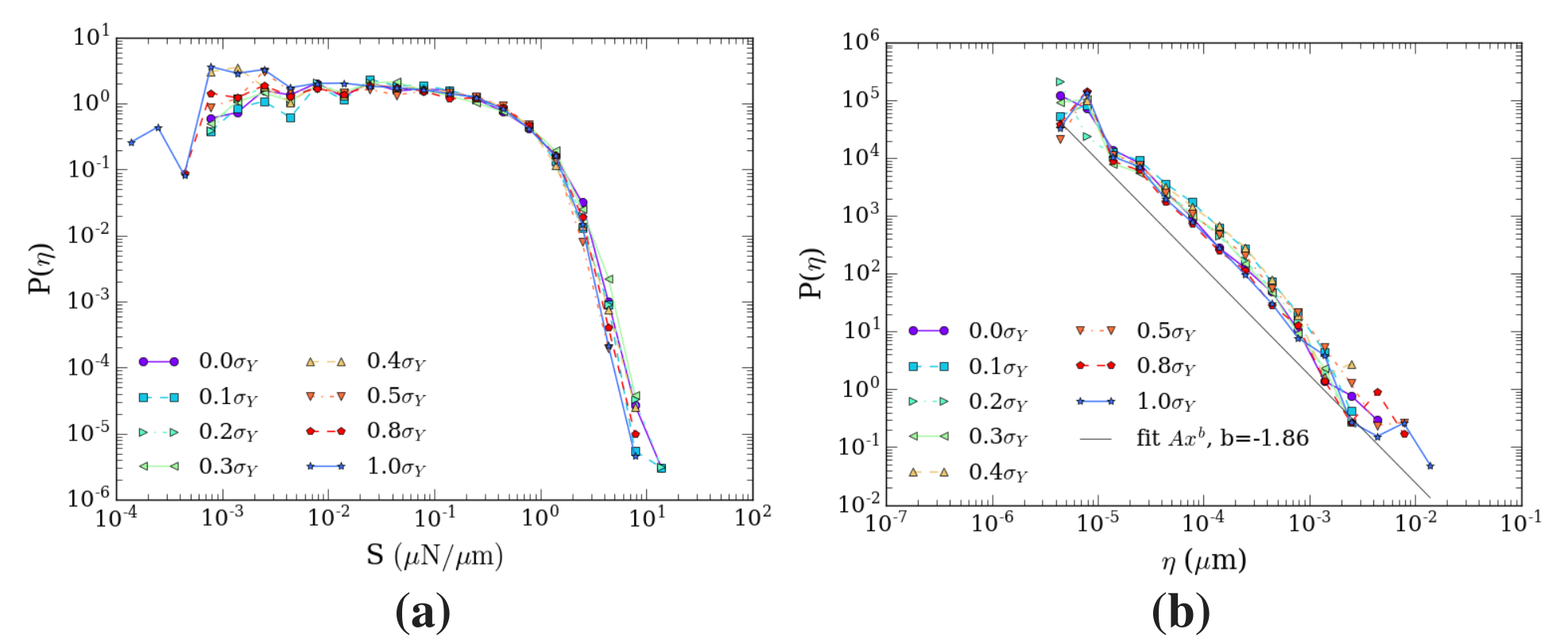}
  \caption{(a) Statistics of indentation force drops, (b) statistics of equivalent indentation depth burst, for a material with source density $\rho_{\rm{nuc}}=75/\mu\rm{m}^2$ and indenter radius $R=1\mu \rm{m}$. The black solid line is a power law fit. $\sigma_{Y}$ is taken as 100 MPa. }
\label{fig:force-drop}
\end{figure}

Under the displacement controlled protocol, the signal of plasticity is a drop of the indentation force  as defined in Fig.~\ref{fig:pop-in-definition}. Statistics of force drops $S$ are shown in Fig.~\ref{fig:force-drop}(a).  Here we barely observe an effect of pre-stress except for very small force drops where pre-stress induced dislocations leads to more noise.  We also define an equivalent indentation depth jump as the distance between the point where the indentation force starts to drop to the point where the force increases back to its initial value (Fig.~\ref{fig:pop-in-definition}). Statistics of the equivalent depth bursts are shown in Fig.~\ref{fig:force-drop}(b).  We see that larger pre-stress can result in larger indentation depth bursts, which is consistent with the statistics for the load controlled protocol shown in  Fig.~\ref{fig:pop-in-statistics}. The power law statistics spans roughly four orders of magnitude with an exponent $-1.86$. 

\section{Discussion and Conclusions}
The main focus of this paper is the effect of pre-stress on the indentation response of single crystals. One important feature observed in section~\ref{hardness-dependence} is that there is a hardness transition (decreasing hardness with increasing pre-stress) depending on pre-stress and indentation depth. This phenomenon is consistent with observations in our recent experimental study~\citep{scripta}.  

Tensile pre-stress basically has two effects on indentation: (i) the stress field reduces the indentation force that is needed to nucleate dislocations, (ii) depending on the magnitude of the pre-stress, it may have introduced pre-existing dislocations prior to indentation. These dislocations got pinned in tension and become mobile in indentation because of the highly concentrated stress field induced by indentation. \cite{Tsui1996} have shown that if the pre-stress is within the elastic regime, there is almost no effect on the indentation hardness measured at a depth of hundreds of nanometers. Consistent with this, the effect of pre-stress found in this paper originates from the pre-stress induced dislocations.
The effect becomes less obvious or disappears at the depth where tension induced dislocations no longer dominate the indentation response, i.e., when dislocations nucleated by indentation dominates. 

\begin{figure}[h!] 
 \centering
 \includegraphics[width=1\textwidth]{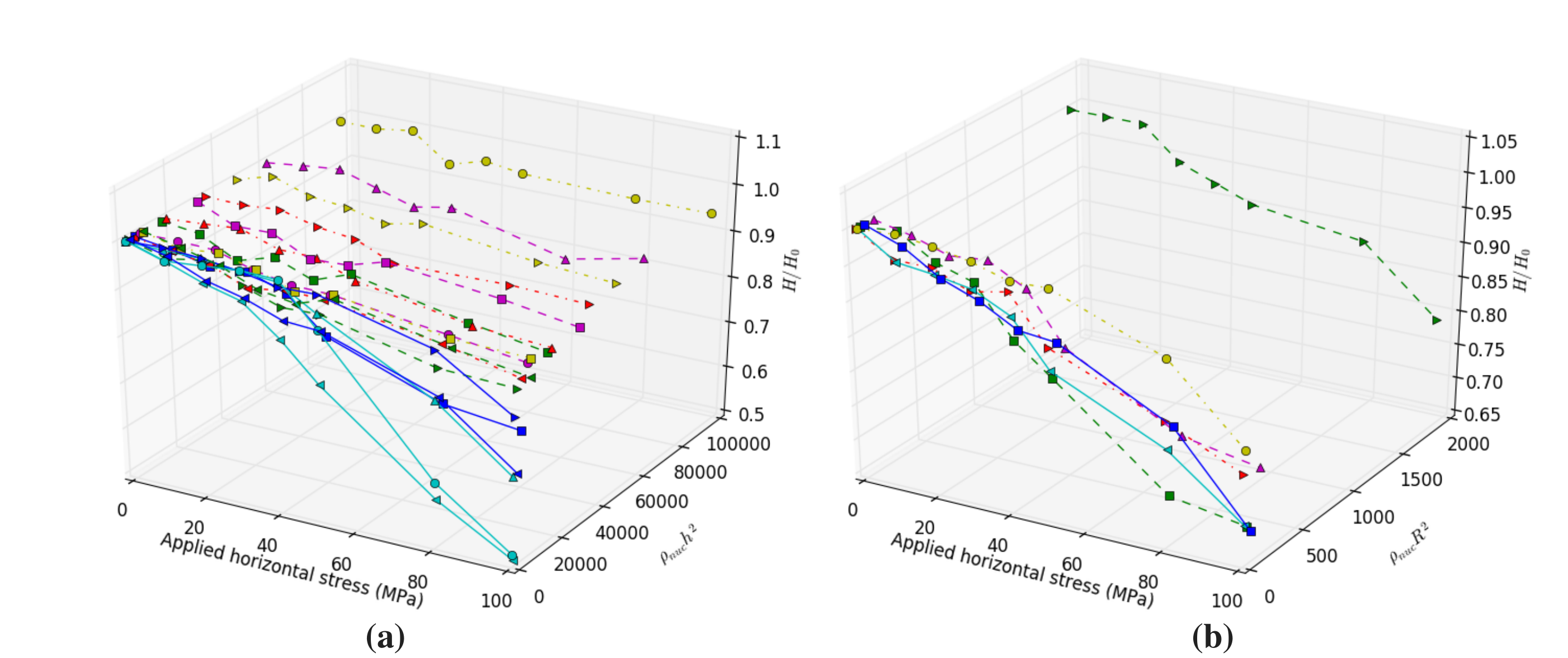}
  \caption{$H/H_{0}$ as the function of pre-stress for different values of the dimensionless quantity $\rho_{\rm{nuc}}h^2$ at constant indented radius $R=1\mu\rm{m}$ in (a) and  versus $\rho_{\rm{nuc}}R^2$ at constant indentation depth $h=10$nm in (b).}
\label{fig:all-hardness}
\end{figure}

For the same dislocation source strength distribution, dislocation nucleation depends on the high stress volume (area, in 2D) and the probability of having dislocation sources in the volume (area). The former depends on indenter geometry and indentation depth, while the latter depends on the source density. It is natural to expect that a smaller indenter (smaller stress volume) and lower dislocation density (lower probability of having sources) shift the transition from pre-existing dislocation-dominated to dislocation nucleation-dominated behavior to larger indentation depths, as observed in sections~\ref{densityeffect} and~\ref{radius}.  Fig.~\ref{fig:all-hardness}(a) shows all computed hardness data for an indenter radius $R=1\mu\rm{m}$ (different indentation depth $h$ and different dislocation source density $\rho_{\rm{nuc}}$) reported in the above sections versus the tensile pre-stress and the dimensionless quantity $\rho_{\rm{nuc}}h^2$. The tensile pre-stress determines the number of dislocations prior to indentation while $\rho_{\rm{nuc}}h^2$ influences dislocation nucleation during indentation. The more dislocations are nucleated during indentation, the weaker the effect of pre-stress induced dislocations, and the transition of hardness fades.  Fig.~\ref{fig:all-hardness}(b) shows, for indentation depth 10nm, normalized hardness as the function of pre-stress and dimensionless quantity $\rho_{\rm{nuc}}R^2$. It shows that increasing $\rho_{\rm{nuc}}R^2$ results in a weaker dependence of hardness transition on the magnitude of the pre-stress. 

In summary, the salient conclusions of this study are: 
\begin{itemize}
\item Larger tensile pre-stress results in a smaller indentation hardness when the indentation depth is small. 
\item The indentation depth that influences the role of pre-stress depends on indenter geometry and material properties (source strength). The effect of pre-stress depends on the competition between pre-stress induced dislocations and indentation nucleated dislocations.
\item Nanoindentation addresses source limited plasticity at small indentation depth and interrogates the strain hardening regime at larger depth. 
\item Statistics for both load controlled and displacement controlled show that pre-stress introduces more {\it noise} and leads to more big events in indentation.  
\end{itemize}     
\section*{Acknowledgement}
This research was supported by the U.S. Department of Energy, Office of Sciences, Basic Energy Sciences, DE-SC0014109. We also acknowledge the use of the Super Computing System (Spruce Knob) at WVU, which are funded in part by the National Science Foundation EPSCoR Research Infrastructure Improvement Cooperative Agreement 1003907, the state of West Virginia (WVEPSCoR via the Higher Education Policy Commission) and WVU. This work also benefited greatly from the facilities and staff of the High Performance Computing Center at the Zernike Institute for Advanced Materials at the University of Groningen, the Netherlands.

\section*{References}

\begin{thebibliography}{57}
\newcommand{\enquote}[1]{``#1''}
\providecommand{\natexlab}[1]{#1}
\providecommand{\url}[1]{\texttt{#1}}
\providecommand{\urlprefix}{URL }
\providecommand{\bibAnnoteFile}[1]{%
  \IfFileExists{#1}{\begin{quotation}\noindent\textsc{Key:} #1\\
  \textsc{Annotation:}\ \input{#1}\end{quotation}}{}}
\providecommand{\bibAnnote}[2]{%
  \begin{quotation}\noindent\textsc{Key:} #1\\
  \textsc{Annotation:}\ #2\end{quotation}}

\bibitem[{Aifantis(1999)}]{aifantis1999strain}
Aifantis, E.C. (1999), \enquote{Strain gradient interpretation of size
  effects.} \emph{International Journal of Fracture}, 95, 299--314.
\input{aifantis1999strain}

\bibitem[{Balint et~al.(2006)Balint, Deshpande, Needleman, and Van~der
  Giessen}]{balint2006discrete}
Balint, D.S., V.S. Deshpande, A.~Needleman, and E.~Van~der Giessen (2006),
  \enquote{Discrete dislocation plasticity analysis of the wedge indentation of
  films.} \emph{Journal of the Mechanics and Physics of Solids}, 54,
  2281--2303.
\input{balint2006discrete}

\bibitem[{Barnoush and Vehoff(2008)}]{Barnoush2008}
Barnoush, A. and H.~Vehoff (2008), \enquote{Hydrogen embrittlement of aluminum
  in aqueous environments examined by in situ electrochemical nanoindentation.}
  \emph{Scripta Materialia}, 58, 747--750.
\input{Barnoush2008}

\bibitem[{Bolshakov et~al.(1996)Bolshakov, Oliver, and Pharr}]{Bolshakov1996}
Bolshakov, A., W.C. Oliver, and G.M. Pharr (1996), \enquote{Influences of
  stress on the measurement of mechanical properties using nanoindentation:
  Part ii. finite element simulations.} \emph{Journal of Materials Research},
  11, 760--768.
\input{Bolshakov1996}

\bibitem[{Bradby et~al.(2004)Bradby, Williams, and Swain}]{Bradby2004}
Bradby, J.E., J.S. Williams, and M.V. Swain (2004), \enquote{Pop-in events
  induced by spherical indentation in compound semiconductors.} \emph{Journal
  of Materials research}, 19, 380--386.
\input{Bradby2004}

\bibitem[{Chang et~al.(2010)Chang, Fivel, Rodney, and
  Verdier}]{fivel-indentation}
Chang, H., M.~Fivel, D.~Rodney, and M.~Verdier (2010), \enquote{Multiscale
  modelling of indentation in fcc metals: From atomic to continuum.}
  \emph{Comptes Rendus Physique}, 11, 285--292.
\input{fivel-indentation}

\bibitem[{Cui et~al.(2014)Cui, Lin, Liu, and Zhuang}]{cui2014theoretical}
Cui, Y., P.~Lin, Z.~Liu, and Z.~Zhuang (2014), \enquote{Theoretical and
  numerical investigations of single arm dislocation source controlled plastic
  flow in fcc micropillars.} \emph{International Journal of Plasticity}, 55,
  279--292.
\input{cui2014theoretical}

\bibitem[{Cui et~al.(2016)Cui, Po, and Ghoniem}]{Cui2016}
Cui, Y., G.~Po, and N.~Ghoniem (2016), \enquote{Controlling strain bursts and
  avalanches at the nano-to micrometer scale.} \emph{Physical review letters},
  117, 155502.
\input{Cui2016}

\bibitem[{Cui et~al.(2017)Cui, Po, and Ghoniem}]{Cui2017}
Cui, Y., G.~Po, and N.~Ghoniem (2017), \enquote{Influence of loading control on
  strain bursts and dislocation avalanches at the nanometer and micrometer
  scale.} \emph{Physical Review B}, 95, 064103.
\input{Cui2017}

\bibitem[{Deshpande et~al.(2005)Deshpande, Needleman, and Van~der
  Giessen}]{Deshpande2005}
Deshpande, V.S., A.~Needleman, and E.~Van~der Giessen (2005),
  \enquote{Plasticity size effects in tension and compression of single
  crystals.} \emph{Journal of the Mechanics and Physics of Solids}, 53,
  2661--2691.
\input{Deshpande2005}

\bibitem[{Dimiduk et~al.(2005)Dimiduk, Uchic, and Parthasarathy}]{Dimiduk2005}
Dimiduk, D.M., M.D. Uchic, and T.A. Parthasarathy (2005),
  \enquote{Size-affected single-slip behavior of pure nickel microcrystals.}
  \emph{Acta Materialia}, 53, 4065--4077.
\input{Dimiduk2005}

\bibitem[{Dimiduk et~al.(2007)Dimiduk, Uchic, Rao, Woodward, and
  Parthasarathy}]{Dimiduk2007}
Dimiduk, D.M., M.D. Uchic, S.I. Rao, C.~Woodward, and T.A. Parthasarathy
  (2007), \enquote{Overview of experiments on microcrystal plasticity in
  fcc-derivative materials: selected challenges for modelling and simulation of
  plasticity.} \emph{Modelling and Simulation in Materials Science and
  Engineering}, 15, 135.
\input{Dimiduk2007}

\bibitem[{Durst et~al.(2006)Durst, Backes, Franke, and G{\"o}ken}]{Durst2006}
Durst, K., B.~Backes, O.~Franke, and M.~G{\"o}ken (2006), \enquote{Indentation
  size effect in metallic materials: Modeling strength from pop-in to
  macroscopic hardness using geometrically necessary dislocations.} \emph{Acta
  Materialia}, 54, 2547--2555.
\input{Durst2006}

\bibitem[{Fleck et~al.(1994)Fleck, Muller, Ashby, and
  Hutchinson}]{Hutchinson-SGP}
Fleck, N.A., G.M. Muller, M.F. Ashby, and J.W. Hutchinson (1994),
  \enquote{Strain gradient plasticity: theory and experiment.} \emph{Acta
  Metallurgica et Materialia}, 42, 475--487.
\input{Hutchinson-SGP}

\bibitem[{Gao et~al.(1999)Gao, Huang, Nix, and Hutchinson}]{SGP-Gao}
Gao, H., Y.~Huang, W.D. Nix, and J.W. Hutchinson (1999),
  \enquote{Mechanism-based strain gradient plasticity—i. theory.}
  \emph{Journal of the Mechanics and Physics of Solids}, 47, 1239--1263.
\input{SGP-Gao}

\bibitem[{Greer and Nix(2006)}]{Greer2006}
Greer, J.R. and W.D. Nix (2006), \enquote{Nanoscale gold pillars strengthened
  through dislocation starvation.} \emph{Physical Review B}, 73, 245410.
\input{Greer2006}

\bibitem[{Greer et~al.(2005)Greer, Oliver, and Nix}]{Greer2005}
Greer, J.R., W.C. Oliver, and W.D. Nix (2005), \enquote{Size dependence of
  mechanical properties of gold at the micron scale in the absence of strain
  gradients.} \emph{Acta Materialia}, 53, 1821--1830.
\input{Greer2005}

\bibitem[{Huang et~al.(2004)Huang, Qu, Hwang, Li, and Gao}]{Huang2004}
Huang, Y., S.~Qu, K.~Hwang, M.~Li, and H.~Gao (2004), \enquote{A conventional
  theory of mechanism-based strain gradient plasticity.} \emph{International
  Journal of Plasticity}, 20, 753--782.
\input{Huang2004}

\bibitem[{Huang et~al.(2000)Huang, Xue, Gao, Nix, and Xia}]{huang2000study}
Huang, Y., Z.~Xue, H.~Gao, W.D. Nix, and Z.C. Xia (2000), \enquote{A study of
  microindentation hardness tests by mechanism-based strain gradient
  plasticity.} \emph{Journal of Materials Research}, 15, 1786--1796.
\input{huang2000study}

\bibitem[{Jang et~al.(2003)Jang, Son, Lee, Choi, and Kwon}]{Jang2003}
Jang, J., D.~Son, Y.~Lee, Y.~Choi, and D.~Kwon (2003), \enquote{Assessing
  welding residual stress in a335 p12 steel welds before and after
  stress-relaxation annealing through instrumented indentation technique.}
  \emph{Scripta Materialia}, 48, 743--748.
\input{Jang2003}

\bibitem[{Lee and Kwon(2004)}]{Lee2004}
Lee, Y. and D.~Kwon (2004), \enquote{Estimation of biaxial surface stress by
  instrumented indentation with sharp indenters.} \emph{Acta Materialia}, 52,
  1555--1563.
\input{Lee2004}

\bibitem[{Li and Bhushan(2002)}]{li2002review}
Li, X. and B.~Bhushan (2002), \enquote{A review of nanoindentation continuous
  stiffness measurement technique and its applications.} \emph{Materials
  characterization}, 48, 11--36.
\input{li2002review}

\bibitem[{Lorenz et~al.(2003)Lorenz, Zeckzer, Hilpert, Grau, Johansen, and
  Leipner}]{Lorenz2003}
Lorenz, D., A.~Zeckzer, U.~Hilpert, P.~Grau, H.~Johansen, and H.S. Leipner
  (2003), \enquote{Pop-in effect as homogeneous nucleation of dislocations
  during nanoindentation.} \emph{Physical Review B}, 67, 172101.
\input{Lorenz2003}

\bibitem[{Maass and Derlet(2017)}]{maass2017micro}
Maass, R. and P.M. Derlet (2017), \enquote{Micro-plasticity and recent insights
  from intermittent and small-scale plasticity.} \emph{arXiv preprint
  arXiv:1704.07297}.
\input{maass2017micro}

\bibitem[{Minor et~al.(2006)Minor, Asif, Shan, Stach, Cyrankowski, Wyrobek, and
  Warren}]{Minor2006}
Minor, A.M., S.S. Asif, Z.~Shan, E.A. Stach, E.~Cyrankowski, T.J. Wyrobek, and
  O.L. Warren (2006), \enquote{A new view of the onset of plasticity during the
  nanoindentation of aluminium.} \emph{Nature materials}, 5, 697--702.
\input{Minor2006}

\bibitem[{Morris et~al.(2011)Morris, Bei, Pharr, and George}]{Morris2011}
Morris, J.R., H.~Bei, G.M. Pharr, and E.P. George (2011), \enquote{Size effects
  and stochastic behavior of nanoindentation pop in.} \emph{Physical review
  letters}, 106, 165502.
\input{Morris2011}

\bibitem[{Ng and Ngan(2008)}]{NG2008}
Ng, K. and A.~Ngan (2008), \enquote{A monte carlo model for the intermittent
  plasticity of micro-pillars.} \emph{Modelling and Simulation in Materials
  Science and Engineering}, 16, 055004.
\input{NG2008}

\bibitem[{Nix and Gao(1998)}]{nix1998indentation}
Nix, W.D. and H.~Gao (1998), \enquote{Indentation size effects in crystalline
  materials: a law for strain gradient plasticity.} \emph{Journal of the
  Mechanics and Physics of Solids}, 46, 411--425.
\input{nix1998indentation}

\bibitem[{Oh et~al.(2009)Oh, Legros, Kiener, and Dehm}]{oh2009situ}
Oh, S., M.~Legros, D.~Kiener, and G.~Dehm (2009), \enquote{In situ observation
  of dislocation nucleation and escape in a submicrometre aluminium single
  crystal.} \emph{Nature materials}, 8, 95--100.
\input{oh2009situ}

\bibitem[{Oyen(2010)}]{oyen2010handbook}
Oyen, M.L. (2010), \emph{Handbook of Nanoindentation: with biological
  applications}. Pan Stanford Publishing.
\input{oyen2010handbook}

\bibitem[{Papanikolaou et~al.(2011)Papanikolaou, Bohn, Sommer, Durin, Zapperi,
  and Sethna}]{Papanikolaou2011}
Papanikolaou, S., F.~Bohn, R.L. Sommer, G.~Durin, S.~Zapperi, and J.P. Sethna
  (2011), \enquote{Universality beyond power laws and the average avalanche
  shape.} \emph{Nature Physics}, 7, 316--320.
\input{Papanikolaou2011}

\bibitem[{Papanikolaou et~al.(2017{\natexlab{a}})Papanikolaou, Cui, and
  Ghoniem}]{papanikolaou2017avalanches}
Papanikolaou, S., Y.~Cui, and N.~Ghoniem (2017{\natexlab{a}}),
  \enquote{Avalanches and plastic flow in crystal plasticity: An overview.}
  \emph{arXiv preprint arXiv:1705.06843}.
\input{papanikolaou2017avalanches}

\bibitem[{Papanikolaou et~al.(2012)Papanikolaou, Dimiduk, Choi, Sethna, Uchic,
  Woodward, and Zapperi}]{Papanikolaou2012}
Papanikolaou, S., D.M. Dimiduk, W.~Choi, J.P. Sethna, M.D. Uchic, C.F.
  Woodward, and S.~Zapperi (2012), \enquote{Quasi-periodic events in crystal
  plasticity and the self-organized avalanche oscillator.} \emph{Nature}, 490,
  517--521.
\input{Papanikolaou2012}

\bibitem[{Papanikolaou et~al.(2017{\natexlab{b}})Papanikolaou, Song, and
  Van~der Giessen}]{Papanikolaou2017}
Papanikolaou, S., H.~Song, and E.~Van~der Giessen (2017{\natexlab{b}}),
  \enquote{Obstacles and sources in dislocation dynamics: Strengthening and
  statistics of abrupt plastic events in nanopillar compression.} \emph{Journal
  of the Mechanics and Physics of Solids}.
\input{Papanikolaou2017}

\bibitem[{Parthasarathy et~al.(2007)Parthasarathy, Rao, Dimiduk, Uchic, and
  Trinkle}]{Parthasarathy2007}
Parthasarathy, T.A., S.I. Rao, D.M. Dimiduk, M.D. Uchic, and D.R. Trinkle
  (2007), \enquote{Contribution to size effect of yield strength from the
  stochastics of dislocation source lengths in finite samples.} \emph{Scripta
  Materialia}, 56, 313--316.
\input{Parthasarathy2007}

\bibitem[{Pharr et~al.(2010)Pharr, Herbert, and Gao}]{pharr2010indentation}
Pharr, G.M., E.G. Herbert, and Y.~Gao (2010), \enquote{The indentation size
  effect: a critical examination of experimental observations and mechanistic
  interpretations.} \emph{Annual Review of Materials Research}, 40, 271--292.
\input{pharr2010indentation}

\bibitem[{Sethna et~al.(2016)Sethna, Bierbaum, Dahmen, Goodrich, Greer, Hayden,
  Kent-Dobias, Lee, Liarte, Ni et~al.}]{sethna2016deformation}
Sethna, J.P., M.K. Bierbaum, K.A. Dahmen, C.P. Goodrich, J.R. Greer, L.X.
  Hayden, J.P. Kent-Dobias, E.D. Lee, D.B. Liarte, X.~Ni, et~al. (2016),
  \enquote{Deformation of crystals: Connections with statistical physics.}
  \emph{arXiv preprint arXiv:1609.05838}.
\input{sethna2016deformation}

\bibitem[{Shim et~al.(2008)Shim, Bei, George, and Pharr}]{Shim2008}
Shim, S., H.~Bei, E.P. George, and G.M. Pharr (2008), \enquote{A different type
  of indentation size effect.} \emph{Scripta Materialia}, 59, 1095--1098.
\input{Shim2008}


\bibitem[{Suresh and Giannakopoulos(1998)}]{Suresh1998}
Suresh, S. and A.E. Giannakopoulos (1998), \enquote{A new method for estimating
  residual stresses by instrumented sharp indentation.} \emph{Acta Materialia},
  46, 5755--5767.
\input{Suresh1998}

\bibitem[{Swadener et~al.(2002)Swadener, George, and
  Pharr}]{swadener2002correlation}
Swadener, J.G., E.P. George, and G.M. Pharr (2002), \enquote{The correlation of
  the indentation size effect measured with indenters of various shapes.}
  \emph{Journal of the Mechanics and Physics of Solids}, 50, 681--694.
\input{swadener2002correlation}

\bibitem[{Swadener et~al.(2001)Swadener, Taljat, and Pharr}]{Swadener2001}
Swadener, J.G., B.~Taljat, and G.M. Pharr (2001), \enquote{Measurement of
  residual stress by load and depth sensing indentation with spherical
  indenters.} \emph{Journal of Materials Research}, 16, 2091--2102.
\input{Swadener2001}

\bibitem[{Tsui et~al.(1996)Tsui, Oliver, and Pharr}]{Tsui1996}
Tsui, T.Y., W.C. Oliver, and G.M. Pharr (1996), \enquote{Influences of stress
  on the measurement of mechanical properties using nanoindentation: Part i.
  experimental studies in an aluminum alloy.} \emph{Journal of Materials
  Research}, 11, 752--759.
\input{Tsui1996}

\bibitem[{Uchic et~al.(2002)Uchic, Dimiduk, Florando, and Nix}]{Uchic03}
Uchic, M.D., D.M. Dimiduk, J.N. Florando, and W.D. Nix (2002),
  \enquote{Exploring specimen size effects in plastic deformation of ni 3 (al,
  ta).} In \emph{MRS Proceedings}, volume 753, BB1--4, Cambridge Univ Press.
\input{Uchic03}

\bibitem[{Uchic et~al.(2004)Uchic, Dimiduk, Florando, and Nix}]{Uchic04}
Uchic, M.D., D.M. Dimiduk, J.N. Florando, and W.D. Nix (2004), \enquote{Sample
  dimensions influence strength and crystal plasticity.} \emph{Science}, 305,
  986--989.
\input{Uchic04}

\bibitem[{Van~der Giessen and Needleman(1995)}]{vandergiessen1995}
Van~der Giessen, E. and A.~Needleman (1995), \enquote{Discrete dislocation
  plasticity: a simple planar model.} \emph{Modelling and Simulation in
  Materials Science and Engineering}, 3, 689.
\input{vandergiessen1995}

\bibitem[{VanLandingham(2003)}]{Vanlandingham2003}
VanLandingham, M.R. (2003), \enquote{Review of instrumented indentation.}
  \emph{Journal of Research of the National Institute of Standards and
  Technology}, 108, 249.
\input{Vanlandingham2003}

\bibitem[{Wang et~al.(2012)Wang, Zhong, Lu, Lu, McDowell, and
  Zhu}]{wang2012size}
Wang, W., Y.~Zhong, K.~Lu, L.~Lu, D.L. McDowell, and T.~Zhu (2012),
  \enquote{Size effects and strength fluctuation in nanoscale plasticity.}
  \emph{Acta Materialia}, 60, 3302--3309.
\input{wang2012size}

\bibitem[{Wei and Hutchinson(1997)}]{wei1997steady}
Wei, Y. and J.W. Hutchinson (1997), \enquote{Steady-state crack growth and work
  of fracture for solids characterized by strain gradient plasticity.}
  \emph{Journal of the Mechanics and Physics of Solids}, 45, 1253--1273.
\input{wei1997steady}

\bibitem[{Widjaja et~al.(2006)Widjaja, Needleman, and Van~der
  Giessen}]{widjaja2006effect}
Widjaja, A., A.~Needleman, and E.~Van~der Giessen (2006), \enquote{The effect
  of indenter shape on sub-micron indentation according to discrete dislocation
  plasticity.} \emph{Modelling and Simulation in Materials Science and
  Engineering}, 15, S121.
\input{widjaja2006effect}

\bibitem[{Widjaja et~al.(2007)Widjaja, Van~der Giessen, Deshpande, and
  Needleman}]{widjaja2007contact}
Widjaja, A., E.~Van~der Giessen, V.S. Deshpande, and A.~Needleman (2007),
  \enquote{Contact area and size effects in discrete dislocation modeling of
  wedge indentation.} \emph{Journal of materials research}, 22, 655--663.
\input{widjaja2007contact}

\bibitem[{Widjaja et~al.(2005)Widjaja, Van~der Giessen, and
  Needleman}]{indentation-Widjaja}
Widjaja, A., E.~Van~der Giessen, and A.~Needleman (2005), \enquote{Discrete
  dislocation modelling of submicron indentation.} \emph{Materials Science and
  Engineering: A}, 400, 456--459.
\input{indentation-Widjaja}

\bibitem[{Xia et~al.(2016)Xia, Gao, Pharr, and Bei}]{Xia2016}
Xia, Y., Y.~Gao, G.M. Pharr, and H.~Bei (2016), \enquote{Single versus
  successive pop-in modes in nanoindentation tests of single crystals.}
  \emph{Journal of Materials Research}, 31, 2065--2075.
\input{Xia2016}

\bibitem[{Xue et~al.(2002)Xue, Huang, and Li}]{particleSGP}
Xue, Z., Y.~Huang, and M.~Li (2002), \enquote{Particle size effect in metallic
  materials: a study by the theory of mechanism-based strain gradient
  plasticity.} \emph{Acta Materialia}, 50, 149--160.
\input{particleSGP}

\bibitem[{Yavas et~al.(2017)Yavas, Song, Hemker, and Papanikolaou}]{scripta}
Yavas, H., H.~Song, K.J. Hemker, and S.~Papanikolaou (2017), \enquote{Detecting
  the plane-tension induced crystal plasticity transition using nanoindentation
  as a probe: Experiments and simulations.} \emph{Submitted}.
\input{scripta}

\end{thebibliography}

\end{document}